\documentclass[aps,prl,twocolumn,superscriptaddress]{revtex4-1}

\usepackage{amsfonts}
\usepackage{amssymb}
\usepackage{amsmath}
\usepackage{upgreek}
\usepackage{color} 
\usepackage{lineno}
\usepackage[percent]{overpic}
\usepackage[export]{adjustbox}
\usepackage{stackengine}
\usepackage{appendix}

\usepackage{graphicx}
\usepackage{hyperref}
\hypersetup{colorlinks=true,
            citecolor=[rgb]{0,0,0.502},
            urlcolor=[rgb]{0,0,0.502},
            linkcolor=[rgb]{0,0,0.502}}

\newcommand*\patchAmsMathEnvironmentForLineno[1]{%
  \expandafter\let\csname old#1\expandafter\endcsname\csname #1\endcsname
  \expandafter\let\csname oldend#1\expandafter\endcsname\csname end#1\endcsname
  \renewenvironment{#1}%
     {\linenomath\csname old#1\endcsname}%
     {\csname oldend#1\endcsname\endlinenomath}}%
\newcommand*\patchBothAmsMathEnvironmentsForLineno[1]{%
  \patchAmsMathEnvironmentForLineno{#1}%
  \patchAmsMathEnvironmentForLineno{#1*}}%
\AtBeginDocument{%
\patchBothAmsMathEnvironmentsForLineno{equation}%
\patchBothAmsMathEnvironmentsForLineno{align}%
\patchBothAmsMathEnvironmentsForLineno{flalign}%
\patchBothAmsMathEnvironmentsForLineno{alignat}%
\patchBothAmsMathEnvironmentsForLineno{gather}%
\patchBothAmsMathEnvironmentsForLineno{multline}%
}

\newcommand{\mytitle}{Ramsey interferometry of  non-Hermitian  quantum impurities  }

\newcommand{\Mainz}{Institut f\"ur Physik, Johannes Gutenberg Universit\"at Mainz, D-55099 Mainz, Germany}
\newcommand{\Harvard}{Department of Physics, Harvard University, Cambridge, Massachusetts 02138, USA}

\begin{document}

\title{\mytitle}

\author{F. Tonielli}
\affiliation{Institut  f\"ur Theoretische Physik, Universit\"at  zu K\"oln, D-50937 Cologne, Germany}
\author{N. Chakraborty}
\affiliation{Centre for Advanced 2D Materials, National University of Singapore, 6 Science Drive 2, 117546, Singapore}
\affiliation{Yale-NUS College, 16 College Avenue West, 138527, Singapore}
\affiliation{Rudolf Peierls Centre for Theoretical Physics, Clarendon Laboratory, Parks Road, Oxford OX1 3PU, UK}
\author{F. Grusdt}
\affiliation{Munich Center for Quantum Science and Technology (MCQST), Schellingstraße 4, 80799 M\"unchen, Germany}
\affiliation{Fakult\"at  f\"ur Physik, Ludwig-Maximilians-Universit\"at, 80799 M\"unchen, Germany}
\author{J. Marino}
\affiliation{\Mainz}
\affiliation{\Harvard}

\def\mean#1{\left< #1 \right>}

\date{\today}

\begin{abstract}

We introduce a  Ramsey pulse scheme which  extracts the  non-Hermitian   Hamiltonian  associated to an arbitrary Lindblad dynamics. 
%
We propose a related protocol to measure via interferometry a generalised Loschmidt echo of a generic state evolving in time with the non-Hermitian Hamiltonian itself, and we apply the scheme to a one-dimensional weakly interacting Bose gas  coupled to a stochastic atomic impurity. The Loschmidt echo is mapped  into a functional integral from which we  calculate the   long-time decohering dynamics   at arbitrary impurity strengths. 
For strong dissipation  we uncover the phenomenology of a quantum many-body Zeno effect: corrections to the decoherence exponent resulting from the impurity self-energy becomes purely imaginary, in contrast  to the regime of small dissipation where they instead   enhance the decay of quantum coherences. 
Our results illustrate the prospects for  experiments employing  Ramsey interferometry to  study  dissipative quantum impurities in condensed matter and cold atoms systems.
\end{abstract}

\maketitle


An  atomic wire subject to localised particle losses, or  a quantum spin chain subject to  on-site dephasing,   are  instances of    dissipative impurity problems. 
The nomenclature is borrowed from the traditional research field of   equilibrium quantum impurities in  many-body systems, which comprises archetypical cases ranging from X-ray edge singularities to  magnetic impurities embedded in fermions~\cite{mahan2013many, affleck2008quantum}. Systems with quantum impurities have  represented important   stepping stones  in understanding  the physics of strongly correlated systems, and   by   adding localised dissipation on an  extended  system, one could similarly  gain insight in the mechanisms intertwining incoherent processes and  quantum correlations in many-body systems.%
%
~The surge of interest in this modern area of research has been ignited by few recent experiments in cold gases: 
 shining an electron beam on a localised spatial region of an atomic BEC of  $^{87}$Rb atoms~\cite{ger08, Ott1, Ott3, Ott4, PhysRevLett.116.235302, patil2014quantum} induces a Zeno effect which dictates  that atom losses  decrease at strong dissipation.  Dissipative impurities can also constitute  a resource in quantum many body engineering, as they are  employed to implement  scanning gate microscopes  of ultra-cold bosons~\cite{lebrat2019quantized,  corman2019quantized}. 

In its conventional formulation,   the quantum Zeno effect  predicts the freezing of the wavefunction when  frequent  measurements exceed a  rate threshold~\cite{facchi2002quantum, kofman1996quantum, kofman2001zeno,kofman2000acceleration, wiseman2009quantum, misra1977zeno, itano1990quantum,PhysRevLett.87.040402}.
The phenomenon extends beyond the theory of quantum measurement, and it comprises the generic arrest of quantum evolution provided by    stochastic fields~\cite{PhysRevA.65.013404}, including the decoupling of a system from its decohering environment by the application of a sequence of fast pulses~\cite{PhysRevA.69.032314}. 
The connection between Zeno effect and many-body physics has been first drawn in the context of quantum circuits where unitary gates and random-in-space and time projective measurements compete, inducing a transition in the entanglement entropy from  volume to  area law when  measurements become frequent~\cite{li2018quantum}.
Primarily motivated by the experiments in Refs.~\cite{Ott3, Ott4, PhysRevLett.116.235302}, the physics of the Zeno effect has also entered the field of  dissipative impurities.
The effect of $1/\omega$ noise on the transport properties of Kane-Fisher barriers~\cite{kane1992transport,kane1992transmission} have been studied with non-equilibrium Luttinger liquids~\cite{dalla2012dynamics}, while   a series of related works have shown that strong local losses can inhibit particles' emission at the Fermi surface~\cite{froml2019fluctuation,froml2019ultracold, wolff2019non}; the interplay of Zeno physics with many-body correlations has the potential to promote lossy mobile impurities into  a novel class of Fermi-polarons~\cite{wasak2019quantum} and the list of examples could continue~\cite{krapivsky2019free, krapivsky2019freeb}.%

In general, the dynamics induced by a Markovian quantum master equation comprise an imaginary or non-Hermitian Hamiltonian (which is quadratic in the Lindblad operator) in combination with a term describing stochastic jumps driven by   quantum noise~\cite{breuer2002theory,weiss2012quantum,gardiner1991quantum}. 
In the following, we  show that a sequence of Ramsey pulses can experimentally decouple these two contributions, and can be employed  to measure via  interferometry a generalised form of  Loschmidt echo, which evolves solely with the  former. This protocol represents  an operative  route to define non-Hermitian Hamiltonians {and it  could bring new experimental insight into communities focusing on this class of hamiltonians; one example is  the field of non-hermitian topology~\cite{esaki2011edge, yao2018non,rudner2009topological,zeuner2015observation,leykam2017edge,yao2018edge,shen2018topological,gong2018topological} which is currently mostly implemented in photonics platforms~\cite{review}}. 

We apply this scheme to the problem of a dissipative impurity in a weakly interacting Bose gas, and we predict, with functional integrals techniques, a  Zeno effect for the decohering exponent of the echo, which is non-perturbative in the impurity strength.
This is at variance with previous works aiming at studying noise averaged Loschmidt echoes in locally dephasing  spin models~\cite{tonielli2019orthogonality,berdanier2019universal}, which require the full Lindblad evolution, and it represents an alternative route to the study of dissipative impurities, since it does not rely on the measurement of transport properties across  noisy barriers.

\begin{figure}[t!]
\includegraphics[width=8.5cm]{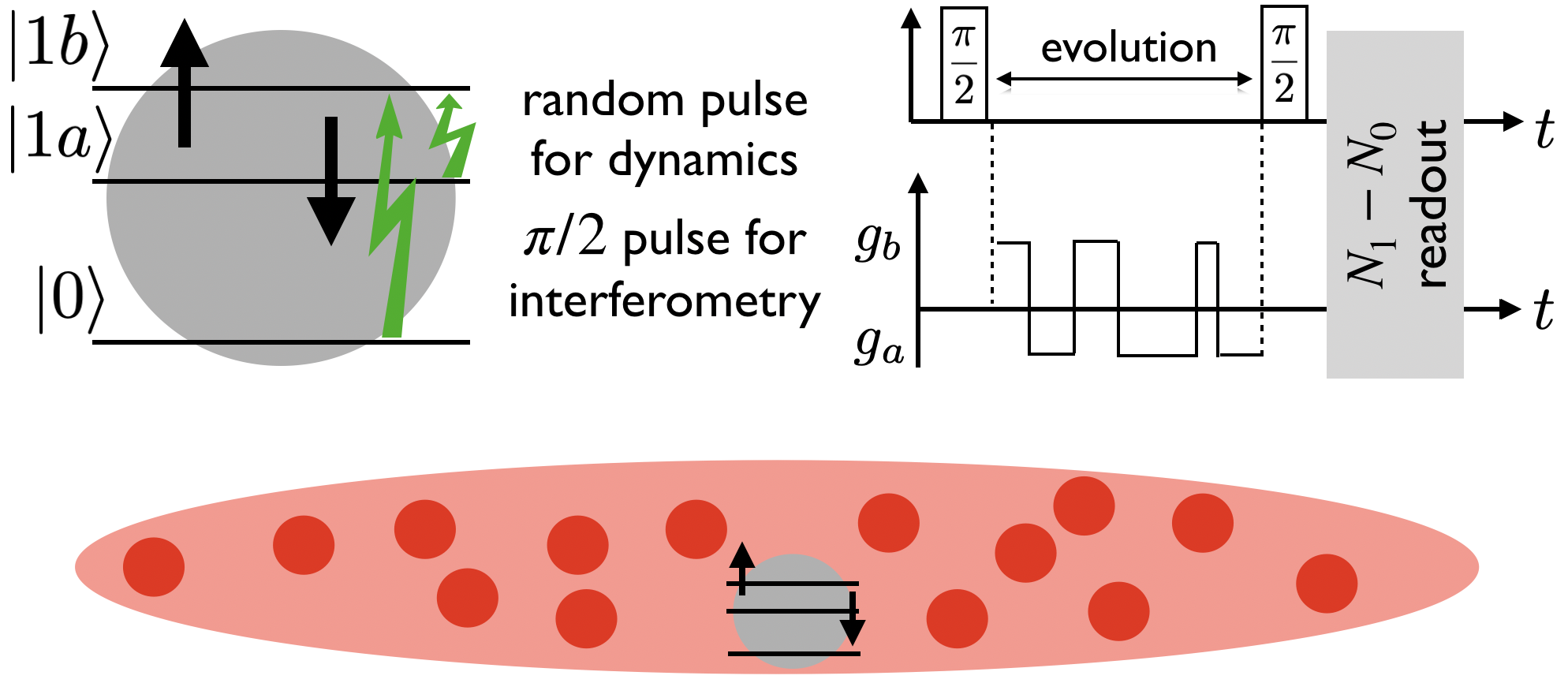}
 \caption{An atomic impurity (grey dot) embedded in a one dimensional wire of cold bosons (in red). The impurity has three internal atomic levels: the level $|0\rangle$ is inert to scattering with the bosonic cloud, and  is used as a control state for   Ramsey interferometry which is employed to measure the Loschmidt echo for the non-Hermitian Hamiltonian associated to stochastic dynamics.  The latter is generated by driving with a sequence of Markovian pulses the levels $|1a\rangle$ and $|1b\rangle$ which are density coupled to the background Bose gas.}
 
\label{fig1}
\end{figure} 

A general Markovian quantum master equation with one Lindblad channel reads
\begin{equation}\label{eq}
i \partial_t\hat{\rho}=\hat{H}^{\,}_{\textsl{eff}}\hat{\rho}-\hat{\rho}\hat{H}^\dag_{\textsl{eff}}+\mathcal{L}_{\textsl{jump}}[\hat{\rho}],
 \end{equation}
with non-Hermitian Hamiltonian and quantum jump term being respectively
\begin{equation}\label{lind}
 \begin{split}
\hat{H}_{\textsl{eff}}=&\hat{H}_0-i\gamma\hat{L}^\dagger\hat{L}\\
\mathcal{L}_{\textsl{jump}}[\hat{\rho}]=&2i\gamma \hat{L}\hat{\rho}\hat{L}^\dagger.
 \end{split}
 \end{equation}
The Ramsey protocol we propose to probe $\hat{H}_{\textsl{eff}}$ relies on the assumption that the strength of the interaction between the system and the environment can be controlled by an additional, discrete degree of freedom of the system. For simplicity, we consider a generic system with (at least) two internal states $|0\rangle$ and $|1\rangle$, the Lindblad dynamics being active only when the system is in the state $|1\rangle$. This is expressed by the following replacement, valid in the extended Hilbert space that includes the internal level:
\begin{align}
\label{statedependent}
    \hat{L}\to\hat{\tilde{L}} = \hat{L}\otimes |1\rangle\langle 1|.
\end{align}
In cold atomic systems, the latter can be realized with spin-dependent interactions with the bath~\cite{chin2010feshbach,Amico_2018} as employed in studies of Bose polarons~\cite{jorgensen2016observation, hu2016bose, yan2019bose}, see Fig.~\ref{fig1}.
 
{The density matrix of the system is prepared in a pure, factorized state $\hat{\tilde\rho}_0 = |\psi\rangle\langle\psi|\otimes|0\rangle\langle0|$. A $\pi/2$ pulse flips the internal state $|0\rangle\to(|0\rangle+|1\rangle)/\sqrt{2}$, and the full density matrix changes accordingly:
\begin{align}
\hat{\tilde\rho}_0\to\hat{\tilde\rho}_1 =   \frac{1}{2}\begin{pmatrix}
     |\psi\rangle\langle\psi| &   |\psi\rangle\langle\psi|  \\
     |\psi\rangle\langle\psi|  &   |\psi\rangle\langle\psi| 
  \end{pmatrix}
\end{align}
At this stage, the crucial observation is that, for a Lindblad operator as in Eq.~\eqref{statedependent}, the quantum jump term acts only on the right-bottom element of the density matrix, i.e., on $\langle 1|\hat{\tilde\rho}(t)|1\rangle$. In order to show that, we consider the action of the Liouvillian on product states of the form 
\begin{align}\label{eq:inista}
     {\tilde{\rho}}=\sum_{n,m=0}^1 p_{nm}  {\rho}\otimes |n\rangle\langle m|.
\end{align}
For   Lindblad operators as in Eq.~\eqref{statedependent}, the action from the left is non-trivial only on components with $n=1$:
\begin{align}
    {\tilde{L}} {\tilde{\rho}}=\big({L}\otimes|1\rangle\langle1|\big){\tilde{\rho}}=\sum_{m=0}^1 p_{1m} {L}{\rho}\otimes |n\rangle\langle m|.
\end{align}
A similar equation holds for the action of ${\tilde{L}}^\dagger$ from the right, with the role of $n$ and $m$ being exchanged; 
the quantum jump term affects therefore only the component of $\tilde{\rho}$ with $n=1, m=1$:
\begin{align}
\mathcal{L}_{jump}[{\tilde\rho}]=2\gamma p_{11}{L}{\rho}{L}^\dagger\otimes|1\rangle\langle 1|.
\end{align}
The action of the full quantum master equation on the components of the density matrix reads therefore
\begin{subequations}
\begin{align}
i\partial_t\tilde\rho_{00}&=p_{00}\big(H_{0}\rho-\rho H_0\big),\\
i\partial_t\tilde\rho_{10}&=p_{10}\big(H_{\textsl{eff}}\rho-\rho H_0\big),\\
i\partial_t\tilde\rho_{01}&=p_{01}\big(H_{0}\rho-\rho H_{\textsl{eff}}^{\dagger}\big),\\
i\partial_t\tilde\rho_{11}&=p_{11}\big(H_{\textsl{eff}}\rho-\rho H_{\textsl{eff}}^{\dagger}+2i\gamma p_{11}L\rho L^{\dagger}\big).
\end{align}
\end{subequations}
The combination of the second $\pi/2$ pulse and of the contrast measurement probes only the non-diagonal components~(8b) and~(8c). In fact, a combination of a $\pi/2$ pulse and a measurement of $\sigma_z$ is equivalent to a measurement of $\sigma_x$, since
\begin{align}
  \langle\sigma_z\rangle=  \text{Tr}[\sigma_z R_{\pi/2}\tilde\rho(t) R^{\dagger}_{\pi/2}] = \text{Tr}[R^{\dagger}_{\pi/2}\sigma_zR_{-\pi/2}\tilde\rho],
\end{align}where 
the matrix of a $\pi/2$-pulse reads $R_{\pi/2}={1}/{\sqrt{2}}(\sigma_z+\sigma_x)$,
 therefore $R_{\pi/2}^{\dagger}\sigma_zR_{\pi/2}=\sigma_x$. In conclusion, denoting by `$\text{tr}$' the trace acting only on the many-body degrees of freedom of the system, we get
\begin{equation}
\begin{split}
\langle \sigma_z \rangle &= \text{Tr}[\sigma_x\tilde\rho(t)]=\text{tr}[\rho_{10}+\rho_{01}]=\\
&=2\text{Re}\,\text{tr}[p_{10}\,e^{iH_0t}e^{-iH_{\textsl{eff}}t}\rho], 
\end{split}
\end{equation}
which yields
 \begin{equation}\label{sigma}
 \langle \hat{\sigma}^z \rangle=\textsl{Re}\left[\langle\psi|e^{i\hat{H}_0t}e^{-i(\hat{H}_0-i\hat{L}^\dag \hat{L} )t} |\psi\rangle\right].
 \end{equation}
Such implementation is inspired from the case of hermitian hamiltonians which we report for completeness in the  Supplemental Material (SM~\cite{Remark1}) (see also   Refs.~\cite{PhysRevX.2.041020, goold2011orthogonality, knap2013probing, cetina2016ultrafast}). 
Such scheme can offer a systematic advantage over methods requiring averages over stochastic realisations, as it occurs in the evaluation of the conventional Loschmidt echo of noise-driven hamiltonians. We finally notice that the derivation starting from Eq.~\eqref{eq:inista} does not require to restrict the initial state to a pure one, and would apply straightforwardly also to mixed states.}

We now particularise our discussion to the setup of an ultracold one-dimensional Bose gas interacting with a localised atomic impurity. The discrete degree of freedom used for interferometry will be an internal state of the atom. To realize Eq.~\eqref{statedependent} {and} the Lindblad dynamics, the atom has a non-interacting internal level $|0\rangle$, and at least two additional levels, labelled as $|1a\rangle$ and $|1b\rangle$, density-coupled to bosons via the interaction Hamiltonian 
\begin{equation}\label{interaction}
\hat{H}_{\textsl{int}}=\sum_{\sigma=a,b}g_\sigma   \hat{n}_\sigma(x_0) \hat{n}_B(x_0),
\end{equation} 
where $x_0$ denotes the position of the atom, $\hat{\psi}_\sigma$ its spinor wavefunction, $\hat{n}_\sigma\equiv \hat{\psi}^\dag_\sigma(x_0)\hat{\psi}_\sigma(x_0)$ the occupation of the $\sigma$ level   and $\hat{n}_B(x_0)$ the bosonic number operator at $x=x_0$; we  assume in the following $x_0=0$. Eq.~\eqref{interaction} describes an interaction whose strength is controlled by the states of the impurity.
The Lindblad dynamics can be engineered by acting within the subspace $\{|1a\rangle,|1b\rangle\}$ with an additional external field, different from the one employed in the Ramsey protocol. With a sequence of $\pi$ pulses flipping between the two states, the coupling can be promoted to a time-dependent quantity $g(t)$, oscillating between $g_a$ and $g_b$, as depicted in \ref{fig1}. For a suitably chosen fast sequence of random pulses, $g(t)$ is a stochastic variable with first and second moments
\begin{align}
\label{moments}
    \langle g(t) \rangle =0,& & \langle g(t)g(t') \rangle =\gamma\delta(t-t'),
\end{align}
where $\langle\cdots\rangle$ denotes the temporal  average over several $\pi$-pulses.
Higher order moments are assumed to be negligible. The conditions on second and higher moments are equivalent to assume that the autocorrelation time of the density operator of the impurity is the smallest scale in the problem.
{
The scheme proposed above, employing a fast sequence of random $\pi$ pulses, can be implemented in ultracold atom systems: Rabi frequencies above 1MHz can be realized in alkali atoms, which leads to fast $\pi$ pulses compared to the typical atomic interaction energies. Meanwhile, couplings to other hyperfine states can be neglected due to large Zeeman splittings on the order of 100 MHz. }

{Our protocol can also be adapted to other experimental platforms, such as superconducting qubit arrays: a given subset of the Hilbert space can be driven with stochastic pulses, while the remaining states can be utilized for measurement purposes. The latter should be prone to dissipation, e.g. as a result of selection rules. Hence our scheme does not necessarily rely on the requirement that the dissipative channels support dark states.} \\

The temporal average over multiple autocorrelation times yields  an equation of motion for the density matrix equal to Eq.~\eqref{eq}, with Lindblad operator $\hat{\tilde{L}}=\hat{n}_B(0)\otimes |1\rangle\langle 1|$. Expanding the Bose field in terms of Bogolyubov excitations, the Lindblad operator and the non-Hermitian Hamiltonian \eqref{lind} become respectively
\begin{align}
   \hat{\tilde{L}}\simeq&-\left(n_0+\sqrt{\frac{n_0}{2\pi}}\int_k V_k (\hat{b}_{k}+\hat{b}^{\dagger}_{k})\right)\otimes |1\rangle\langle 1|
\end{align}
and
\begin{align}
\label{eq:Heff}
    \hat{H}_{\textsl{eff}}=&\,\hat{H}_{0}-i\gamma\hat{L}^{\dagger}\hat{L} =\\
=&\int_k~\omega_{k}\hat{b}^{\dagger}_{k}\hat b_{k} - ig(\rho^{2}+\tilde{\Lambda}_{\tau})\notag\\
-& 2ig\rho\int_kV_{k}(\hat{b}_{k}+\hat{b}^{\dagger}_{-k})\notag \\
  -&ig \int_{kq} V_{k}V_{q}\big(2\hat{b}^{\dagger}_{k}\hat b_{q}+\hat{b}^{\dagger}_{k}\hat b^{\dagger}_{-q}+\hat{b}_{k}\hat b_{q}\big).\notag
\end{align}
where we have defined $\int_k\equiv \int^{\Lambda_\tau}_{-\Lambda_\tau}dk$.
The cut-off $\Lambda_\tau$ is a consequence of the Markov   approximation (cf. Eq.~\eqref{moments}) on  the statistics of the $\pi$-pulses, which are assumed to evolve on the shortest time scale $\tau$ in the model. However, this assumption is no longer valid when the dispersion relation $\omega_k$ enters the particle-like regime and momenta are of the order of $k\simeq \Lambda_\tau\propto \sqrt{1/\tau}$, thus requiring to cut off momentum modes beyond this UV scale. %
The parameter $\tilde{\Lambda}_{\tau}$   in Eq.~\eqref{eq:Heff} comes from the normal ordering of $\hat{H}_{\textsl{eff}}$, and  it is related to the  cutoff $\Lambda_\tau$
 via $\tilde{\Lambda}_{\tau}=\sqrt{2+\Lambda_\tau^2}-\sqrt{2}$.  

In the  expression~\eqref{eq:Heff}, $\hat b_{k}$ are the Bogolyubov annihilation operators in the BEC, $n_0$ is the density of the condensate, and we have defined
\begin{equation}\label{bogo}
\begin{split}
V_{k} =& \left(\frac{k^{2}}{2+k^{2}}\right)^{1/4}, \quad\omega_{k} = |k|\sqrt{1+k^{2}/2},\\
\rho = &\sqrt{2\pi n_{0}}, \quad g=\frac{\gamma n_{0}}{4\pi},
\end{split}
\end{equation}
where $g$ expresses the dissipation strength in $\hat{H}_{\textsl{eff}}$ and replaces the microscopic coupling constant (we have used units  $\hbar=c=m=\xi=1$, where $c$ is the speed of sound, $\xi$ the healing length, $m$ the mass of the bosons). 

The contrast \eqref{sigma} and the related Loschmidt amplitude $\mathcal{G}(t)$ are now expressed in terms of a functional integral with fixed boundary conditions in time, following  the standard coherent state Trotter decomposition. The derivation follows Ref.~\cite{altland2010condensed}, and it is discussed in detail in the Supplemental Material. Specifically, from Eq.~\eqref{eq:Heff} we find
\begin{align}
\label{eq:funcint}
\mathcal{G}(t)&\equiv \langle 0|e^{-it\hat{H}_{eff}}|0\rangle = \notag \\ &=\int_{b_{k}(0)=0}^{b_{k}(t)=0} DbDb^{*} \exp\Bigg(i\int_{0}^{t}ds\bigg[\int_k\mathcal{A}_k\bigg]\Bigg),
\end{align}
where
\begin{align}
\label{eq:funcint}
\mathcal{A}_k\equiv b^{*}_{k}(s) \big(i\partial_{s}-\omega_{k}\big)b^{\,}_{k}(s)+i\gamma L^{*}(s)L(s).
\end{align}

 We remark that the functional integral formula~\eqref{eq:funcint} is suited to describe the outcome of the interferometric measurement discussed above for any choice of Lindblad operator $\hat{L}$, which can be local or extended in space. 
Eq.~\eqref{eq:funcint} is analogous to a Matsubara functional integral in imaginary time, as it can be readily seen from the similarity between the time evolution operator, $\exp{(-it\hat{H})}$, and the Boltzmann weight,  $\exp{(-\beta\hat{H})}$; accordingly, we define the \emph{real time} Matsubara frequencies $\omega_{n}=2\pi n/t$, with $n\in\mathbb{Z}$. Implementing the boundary conditions requires however an additional Lagrange multiplier, as discussed in the Supplemental Material.
 
\begin{figure}[t!]
\includegraphics[width=6.5cm]{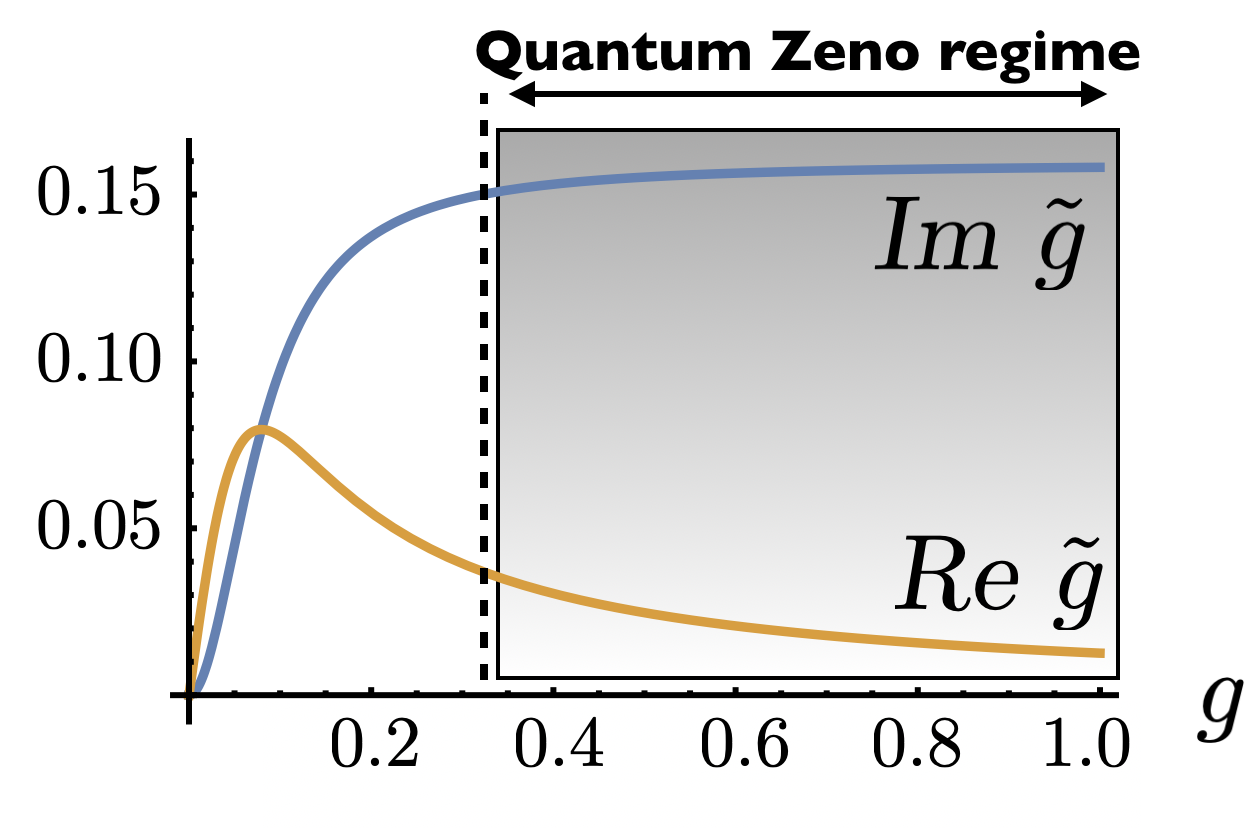}
 \caption{Plot of the real  and imaginary parts of the renormalised impurity strength, $\tilde{g}$, as a function of the bare dissipation strength $g$. For $g\gg1$ the real part  vanishes, while the imaginary part reaches an asymptote at $1/2\pi$, indicating the onset of a quantum Zeno regime.} 
\label{fig2}
\end{figure}

 The {bare}, $G^{0}_{k,n}$, and {impurity dressed}, $G^{eff}_{kq,n}$,  Matsubara Green's functions can be derived from Eq.~\eqref{eq:funcint} after some manipulations which yield (see Supplemental Material for the details of the calculations) 
\begin{subequations}
\begin{align}
G^{0}_{k,n} &= \frac{2V_{k}^{2}\omega^{\,}_{k}}{\omega_{n}^{2}-\omega_{k}^{2}},\\
\label{eq:impurityGF}
G^{eff}_{kq,n} &= G^{0}_{k,n}\delta^{\,}_{k,q} - \frac{2ig}{1+2ig\int_{p}G^{0}_{p,n}}G^{0}_{k,n}G^{0}_{q,n}.
\end{align}
\end{subequations} 
Before proceeding further, we observe that Eq.~\eqref{eq:impurityGF} carries a crucial information on the perturbative expansion of $\log\mathcal{G}(t)$ in powers of the coupling $g$: all corrections corresponding to a dressing of the Green's functions can be resummed and expressed in terms of a \emph{renormalized} coupling strength
\begin{align}\label{dress}
\tilde g \equiv \frac{2g}{1+2ig\int_{p}G^{0}_{p,0}} = \frac{2g}{1-4\pi i g}.
\end{align}
This parameter is { small} since $|\tilde g|_{max}=1/2\pi\simeq 0.16$ (cf. Fig.~\ref{fig2}); it is therefore convenient to develop an expansion of $\log\mathcal{G}$ in powers of $\tilde g$. 
The functional integral \eqref{eq:funcint} can be now evaluated~(see Supplemental Material), obtaining the following exact expression of the Loschmidt amplitude
\begin{equation}
\label{eq:logG}
\begin{split}
&\log\mathcal{G}(t) = -g(\rho^{2}+\tilde{\Lambda}_{\tau})t\\
&+2ig^{2}\rho^{2}t\left(\int_{k,q}G^{eff}_{kq,0}+\text{tr}\Big[G^{eff}_{0}\big(\sum_{n}G^{eff}_{n}\big)^{-1}G^{eff}_{0}\Big]\right)\\
&+\frac{1}{2}\,\sum_{n}\text{tr}\log\left[(G^{0}_{n})^{-1}G^{eff}_{n}\right]\\&-\frac{1}{2}\,\text{tr}\log\Big[\big(\sum_{n}G^{0}_{n}\big)^{-1}\sum_{n}G^{eff}_{n}\Big].\\
\end{split}
\end{equation}

In Eq.~\eqref{eq:logG} matrices act only in momentum space, and $G_{0}$ is a shorthand for $G_{n=0}$; correspondingly, traces run only over momenta. 
We can observe here the role played by the renormalized coupling in the analytic expression. The first term in a naive perturbation theory corresponds to the first line, i.e., it is obtained by replacing $G^{eff}\to G^0$. Crucially, almost all corrections to naive perturbation theory are small, and under perturbative control even at strong coupling, since they can be resummed and expressed in terms of $\tilde g$, as manifested by the presence of the dressed Green's functions in Eq.~\eqref{eq:logG}. The only possibly relevant contribution to the naive perturbation theory comes from the second line, that also contains the bare coupling $g^2$: the leading term at long times can be evaluated exactly, and the sum of first and second line yields the Loschmidt echo
\begin{align}\label{echo}
\log\left(\mathcal{G}(t)/\mathcal{G}_\tau(t)    \right) \simeq - \tilde g\rho^{2}t/2,
\end{align}
in terms of the  amplitude $\mathcal{G}_\tau(t)\equiv \exp(-g\tilde{\Lambda}_{\tau}t)$, which can be controlled by shaping the noise profile. 
The right hand side of   Eq.~\eqref{echo} represents non-perturbative corrections to the leading decoherence damping, $\mathcal{G}_\tau(t)$, expected in general for a stochastic scatterer embedded in an otherwise coherent medium.
Nevertheless, the renormalized coupling $\tilde g$, which is real for small values of the bare coupling $g$,  becomes purely imaginary at strong bare coupling, $\tilde{g}\simeq i(2\pi)^{-1}$, as illustrated in  Fig.~\ref{fig2}. 
The fact that $\tilde{g}$ is imaginary for large dissipative strengths, indicates that the rate decay function of the Loschmidt echo will be entirely dominated by the bare decay exponent~$\propto g\tilde\Lambda$. The occurrence that all  higher order corrections to decoherence are neutralised and resum to an imaginary exponent is an incarnation of the Zeno effect: for strong dissipation (large $g$), the incoherent scatterer perfectly reflects  bosons which impinge upon it, and its only effect is to imprint a phase shift on reflected wavefunctions (see for related ideas the cold-atoms experiment in Ref.~\cite{syassen2008strong}). 

{Conversely, in the conventional case of the Loschmidt echo of a decohering qubit coupled to an extended quantum system (e.g. a quantum spin chain), one expects a rate of decay which monotonously grows with the system bath coupling or with dissipation strength, see for instance Ref.~\cite{rossini2007decoherence}. }\\

In conclusion, we have proposed how to measure, via Ramsey interferometry, the Loschmidt amplitude of an effective non-Hermitian Hamiltonian associated to a Lindbladian. The onset of a many-body Zeno effect can be directly probed by the readout of $\langle \sigma^z \rangle$ without resorting to measurements of transport properties or to probing unequal time correlation functions. We have demonstrated through Eqs.~\eqref{eq:impurityGF} and~\eqref{dress} that unitarity is restored for quasi-particle dynamics at strong dissipation strength, while Eq.~\eqref{eq:logG} and Eq.~\eqref{echo} show that, in spite of the onset of the Zeno effect, a  damping persists and becomes dominant at strong coupling. 
It would be interesting to study, in the future, whether the Zeno effects can manifestly similarly in the interferometric properties of other systems, or whether its imprint on the Loschmidt echo is inherently non-universal. 

Our results pave the way for a number of further   exploratory directions. First of all, it would be natural to study extension of our calculations in the case of a  mobile impurity in view of   recent connections between polarons and Zeno physics~\cite{wasak2019quantum}. 
Furthermore, the approach developed for extracting the leading decay rate of $\mathcal{G}(t)$ is completely general, and it could be, for instance, extended to more realistic dissipative impurities by taking into account the spatial profile of the impurity wavefunction or the correlation time of the noise. 
 Finally, the short-time pattern of the generalised Loschmidt echo could serve as a mean to characterise dynamical quantum phase transitions of non-Hermitian systems \cite{Heyl2013, Heyl2018}. We also foresee the possibility of applying concepts developed for the study of dynamical topological phenomena \cite{Heyl2018,Budich2016} to the more recent field of non-Hermitian topology \cite{bergholtz2019exceptional, SatoUeda}, with direct access to the echo of physically realizable non-Hermitian Hamiltonians defined via  the protocol discussed here.
Since  out-of-time order correlations can be measured via Ramsey interferometry~\cite{ yao2016interferometric}, we also foresee in the future an extension our results in the direction of probing scrambling in non-hermitian quantum systems.

 \begin{acknowledgments} F.T. thanks C. Mordini for useful discussions.
 F.G. acknowledges support by the Deutsche Forschungsgemeinschaft (DFG, German Research Foundation) under Germany's Excellence Strategy -- EXC-2111 --  390814868. 
 J.M. was supported by the European Union's Horizon 2020 research and innovation programme under the Marie Sklodowska-Curie grant agreement No 745608 (QUAKE4PRELIMAT). 
\end{acknowledgments}

\bibliography{impurity_lib}

 \newpage
 
 \begin{widetext}

 \section{\Large{Supplemental Material}}

\section{Ramsey interferometry for Hermitian Hamiltonians}

The same interferometric protocol discussed in the main text can be applied to  Hermitian Hamiltonians. 
We consider the state-dependent interaction encoded in 
\begin{equation}\label{ham}
\tilde{H}=H+|1\rangle\langle 1|\otimes V
\end{equation}

and prepare the  initial state in the tensor product of the many-body wave function $|\psi\rangle$ with the control state $|0\rangle$ of the auxiliary spin

\begin{equation}
|\varphi\rangle=|\psi\rangle\otimes|0\rangle.
\end{equation}

After a $\pi/2$ pulse this state is mapped into the superposition

\begin{equation}
|\varphi'\rangle=|\psi\rangle\otimes\frac{|0\rangle+|1\rangle}{\sqrt{2}},
\end{equation}
whose time evolution reads

\begin{equation}\label{state}
|\varphi'\rangle(t)= e^{-iHt}|\psi\rangle\otimes\frac{|0\rangle}{\sqrt{2}}+e^{-i(H+V)t}|\psi\rangle\otimes\frac{|1\rangle}{\sqrt{2}},
\end{equation}
because of the conditioned activation of the perturbation $V$ upon occupation of the level $|1\rangle$ in the Hamiltonian~\eqref{ham}.
By letting the system evolve for the time $t$ and applying a second $\pi/2$ pulse to the state~\eqref{state}, we find

\begin{equation}
|\varphi'\rangle(t)= e^{-iHt}|\psi\rangle\otimes\frac{|0\rangle+|1\rangle}{\sqrt{2}}+e^{-i(H+V)t}|\psi\rangle\otimes\frac{|0\rangle-|1\rangle}{\sqrt{2}}.
\end{equation}
Measuring the imbalance between the $|1\rangle$ and $|0\rangle$ states, we recover the standard Loschmidt amplitude

\begin{equation}
\langle \sigma^z\rangle=\textsl{Re}\langle\psi|   e^{iHt}e^{-i(H+V)t}|\psi\rangle.
\end{equation}

\section{Matsubara functional integral}

We express $\mathcal{G}(t)$   it in terms of a functional integral with {fixed boundary conditions}. For illustrative purposes, we reproduce here such derivation in the case of a single bosonic mode with Hamiltonian 
\begin{align}
\hat{h}=\hat{b}^{\dagger}\big(\omega_{0}-ig\big)\hat{b}-2ig\rho(\hat{b}+\hat{b}^{\dagger})- ig(\rho^{2}+\tilde\Lambda).
\end{align}
The precise form of the above Hamiltonian is chosen in analogy to Eq.~(9) in the main text.

We begin by identifying the proper boundary conditions which should be imposed on the bosonic fields:
\begin{align}
&\mathcal{G}(t)= \langle 0|e^{-it\hat{h}}|0\rangle =\notag\\
&\quad= \int d^{2}\alpha_{0}\,e^{-|\alpha_{0}|^{2}}\langle \alpha_{0}|e^{-it\hat{h}}|\alpha_{0}\rangle\delta(\alpha_{0}-0), \label{eq:constraint}
\intertext{and}
&\langle \alpha_{0}|e^{-it\hat{h}}|\alpha_{0}\rangle=\int d^{2}\alpha_{N}\langle \alpha_{N}|e^{-it\hat{h}}|\alpha_{0}\rangle\delta(\alpha_{0}-\alpha_{N}). \label{eq:periodicity}
\end{align}

Eq.~\eqref{eq:periodicity} is a periodicity condition $\alpha(0)=\alpha(t)$ and it can be taken into account following the formalism for finite temperature functional integrals, i.e. by introducing  Fourier-transformed fields to take advantage of periodicity in time.

We start by  trading the $\delta$-constraint in Eq.~\eqref{eq:constraint} for an integral over an auxiliary variable
\begin{align}
\label{eq:auxfield}
\delta(\alpha_{0}) = \int d^{2}j\,e^{i(j^{*}\alpha_{0}+\alpha_{0}^{*}j)}.
\end{align}
The exponent is a \emph{boundary term in time}, as it depends only on the field at $t=0$.
Having fixed the boundary term, we note that $\langle \alpha_{N}|e^{-it\hat{h}}|\alpha_{0}\rangle\big|_{\alpha_{0}=\alpha_{N}}$ is  the starting point of the derivation of the Matsubara functional integral (see Ref.~[36]). The only difference with respect to the conventional case   is the position $H\to ih$, and the integration over the auxiliary field as in Eq.~\eqref{eq:auxfield}. The real time functional integral reads in this case 
\begin{align}
\label{eq:FuncInt}
\mathcal{G}(t)& = \int D(\alpha,\alpha^{*})\,dj\, e^{-g(\rho^{2}+\tilde\Lambda)t+iS + i(j^{*}\alpha(0) + j\alpha^{*}(0))}, \\
S&=\int_{0}^{t}ds\big[\alpha^{*}\big(i\partial_{s}-\omega_{0}+ig\big)\alpha+2ig\rho\big(\alpha+\alpha^{*}\big)\big].\notag
\end{align}
Since such  integral is in general not convergent,  one  should   consider the ratio of Eq.~\eqref{eq:FuncInt} with the same expression without the impurity,   $\mathcal{G}_{0}(t)=\langle0|e^{-it\hat{H}_{0}}|0\rangle$. 

The Fourier transform is defined as
\begin{subequations}
\begin{align}
\alpha_{n}=\frac{1}{\sqrt t}\int_{0}^{t}ds\,\alpha(s)e^{i\omega_{n}s},& &\omega_{n}=\frac{2\pi n}{t},\\
\alpha(s)=\frac{1}{\sqrt t}\sum_{n\in\mathbb{Z}}\alpha_{n}e^{-i\omega_{n}s},& &
\end{align}
\end{subequations}
where we have introduced the  {real time Matsubara frequencies} $\omega_{n}$, defined in the main text. In the following, indices will always denote frequency and momentum arguments and continuous arguments will denote time and space variables.
The Fourier transformed action and boundary term read then
\begin{subequations}
\label{eq:ActionsToy}
\begin{align}
\label{eq:MatsubaraActionToy}
S=&\sum_{n}\alpha_{n}^{*}\big(\omega_{n}e^{-i\omega_{n}\delta}-\omega_{0}+ig\big)\alpha_{n}+2ig\rho\sqrt{t}(\alpha_{0}+\alpha^{*}_{0}),\\
\label{eq:boundaryActionToy}
S_{b.}=&\sum_{n}\big[j^{*}\alpha_{n}+j\alpha_{n}^{*}\big],
\end{align}
\end{subequations}
where in Eq.~\eqref{eq:boundaryActionToy} we absorbed a factor  $\sqrt{t}$ in a redefinition of the Lagrange multiplier. The infinitesimal quantity $\delta\to0^+$ in Eq.~\eqref{eq:MatsubaraActionToy} is a positive infinitesimal, it is needed to regularize Matsubara sums, and it comes  from the Trotter decomposition (see Ref.~[36]). We will  label in the following for brevity
\begin{align}
\omega_{n}e^{-i\omega_{n}\delta}\equiv z_{n}.
\end{align}

\section{Green's functions of the model and dressed coupling} We now consider the many-body version of our problem, restore momentum arguments and discuss the relevant Green's functions. 
Relabeling $\alpha\to b$, Eqns.~\eqref{eq:ActionsToy} become
\begin{subequations}
\label{eq:Actions}
\begin{align}
S=&\sum_{n}\int_{k,q}\Big[b^{*}_{n,k}(\omega_{n}-\omega_{k})\delta_{kq}b_{n,q}\\
&+igV_{k}V_{q}(2b^{*}_{n,k}b_{n,q}+b^{*}_{n,k}b^{*}_{-n,-q}+b_{-n,-k}b_{n,q})\Big]\notag\\
&+2ig\rho\sqrt{t}\int_{k}V_{k}(b_{0,k}+b^{*}_{0,-k}),\notag\\
S_{b.}=&\sum_{n}\int_{k}\big[j^{*}_{k}b_{n,k}+j_{k}b_{n,k}^{*}\big]
\end{align}
\end{subequations}

We exploit the structure of the interaction by introducing the bosonic scalar field
\begin{align}
\phi_{n,k}= V_{k}(b_{n,k}+b^{*}_{-n,-k}),
\end{align}
which is the real part of the complex field $b$ up to a pre-factor. The interaction with the impurity becomes simple, and the total action reads schematically 
\begin{align}
\label{eq:ActionScalar}
S[\phi,\tilde{j}]=&\frac{1}{2}\sum_{n}\int_{k,q}\phi_{-n,-q}\big[(G^{0}_{n,k})^{-1}\delta_{kq}+2ig\big]\phi_{n,k}\,\\
&+2ig\rho\sqrt{t}\int_{k}\phi_{0,k}+\sum_{n}\int_{k}\tilde{j}_{-k}\phi_{n,k}.\notag
\end{align}
This action for $\phi$ can either be derived by writing the complex field $b$ in terms of real and imaginary parts, and by integrating out the latter, or can be derived   by noticing that the Green's functions $G^{0}_{n,k}$ of the scalar field can be computed from those of the complex field.
In the following, we will choose the latter approach.

From Eq.~\eqref{eq:Actions}, we find the complex boson Green's function
\begin{align}
-i\langle b_{n,k}b^{*}_{n,q}\rangle = \frac{1}{z_{n}-\omega_{k}}\delta_{kq},
\end{align}
from which the bare scalar Green's function reads
\begin{align}
\label{eq:bareGF}
G^{0}_{n,k}&=-i\langle \phi_{n,k}\phi_{-n,-k}\rangle = \frac{V^{2}_{k}}{z_{n}-\omega_{k}}+\frac{V^{2}_{k}}{z_{-n}-\omega_{k}}\notag\\
&=V_{k}^{2}\frac{-2\omega_{k}+z_{n}+z_{-n}}{\omega_{k}^{2}-\omega_{n}^{2}-\omega_{k}(z_{n}+z_{-n})}\notag.
\end{align}
From $z_{n}=\omega_{n}e^{-i\delta\omega_{n}}$ and from $\omega_{-n}=-\omega_{n}$ one sees that the sum $z_{n}+z_{-n}$ is a negative infinitesimal $\simeq -2i\delta\omega_{n}^{2}$. We neglect such infinitesimal at the numerator since it does not influence the location of poles in the complex plane. Its influence in the denominator is to shift its zeros away from the real axis. The denominator equals $\omega_{k}^{2}-\omega_{n}^{2}(1-2i\delta\omega_{k})$, and, to first order in $\delta$, it vanishes for $\omega_{n}\simeq\pm (\omega_{k}+i\delta \omega^{2}_{k})$. We will label them in the following as $\omega_{k}^{+}\equiv \omega_{k}+i\delta$. 
We therefore obtain for the bare Green's function
\begin{align}
G^{0}_{n,k}=\frac{2V_{k}^{2}\omega_{k}}{\omega_{n}^{2}-(\omega^{+}_{k})^{2}}.
\end{align}
\subsection{Impurity dressed Green's function}
The {impurity dressed} Green's function is obtained after inversion of the quadratic kernel in Eq.~\eqref{eq:ActionScalar}. The inversion is trivial in Matsubara frequency indices (the kernel is diagonal) but not in momentum space. To make progress, we write the Dyson equation for each component of the Green's function (we suppress the index $n$ in the following few lines, and restore it at the  end of the calculation):
\begin{align*}
G^{\textsl{eff}} = [(G^{0})^{-1}+V]^{-1}\quad \Rightarrow\quad G^{\textsl{eff}}=G^{0}-G^{0}VG^{\textsl{eff}}.
\end{align*}
Since $V_{p_{1}p_{2}}=2ig$ does not depend on momentum indices, we find
\begin{align}
G^{\textsl{eff}}_{kq} &= G^{0}_{k}\delta_{kq} - \int_{p_{1}p_{2}}G^{0}_{k}\delta_{kp_{1}}\cdot 2ig\cdot G^{\textsl{eff}}_{p_{2}q}\notag\\
&=G^{0}_{k}\delta_{kq} - 2ig G^{0}_{k}\int_{p}G^{\textsl{eff}}_{pq}\notag\\
\Rightarrow \int_{k}G^{\textsl{eff}}_{kq} &= G^{0}_{q} - 2ig\int_{k}G^{0}_{k}\int_{p}G^{\textsl{eff}}_{pq}\notag\\
\Rightarrow \int_{k}G^{\textsl{eff}}_{kq} &=\frac{1}{1+2ig\int_{p}G^{0}_{p}}G^{0}_{q}.\notag\\
\label{eq:impurityGF}
\Rightarrow G^{\textsl{eff}}_{n,kq} &= G^{0}_{n,k}\delta_{kq} - \underbrace{\frac{2ig}{1+2ig\int_{p}G^{0}_{n,p}}}_{\tilde{g}_{n}}G^{0}_{n,k}G^{0}_{n,q}.
\end{align}
To gain further insight, we perform some algebra in  the denominator of Eq.~\eqref{eq:impurityGF}:
\begin{align}
G^{0}_{n,p}&=G^{0}_{0,p}+\Delta G^{0}_{n,p} = \frac{-2V_{p}^{2}}{\omega_{p}}+\frac{2V_{p}^{2}}{\omega_{p}}\,\frac{\omega_{n}^{2}}{\omega_{n}^{2}-\omega_{p}^{2}},\notag\\
\frac{2V_{p}^{2}}{\omega_{p}}&=\frac{\sqrt{2}}{1+p^{2}/2},\quad \int_{p}G^{0}_{0,p} = -2\pi,\\
\int_{p}\Delta &G^{0}_{n,p}\equiv\zeta_{n}=2\pi-\sqrt{2}\pi\frac{\sqrt{f(\omega_n)+1}-i\sqrt{f(\omega_n)-1}}{f(\omega_n}
,\notag\\
f(\omega_{n})&=\sqrt{1+2\omega_{n}^{2}}.\notag
\end{align}
In the first equation we split $\int_{p}G_{n,p}$ into a constant and a $n$ dependent term. The integrand is bounded for $n\to\infty$ and thus does not alter any convergence property  of Matsubara sums. We therefore write the factor $\tilde{g}_{n}$ in Eq.~\eqref{eq:impurityGF} as
\begin{subequations}
\begin{align}
\tilde{g}_{n}&=\frac{2ig}{1-4\pi ig +2ig\zeta_{n}}\equiv\frac{\tilde g}{1+\tilde{g}\zeta_{n}},\\
\tilde{g} &\equiv \frac{2ig}{1-4\pi ig},
\end{align}
\end{subequations}
which is one of the key results in the main text.

\subsection{Integrating over bosons and  auxiliary fields}

We rewrite the bosonic action \eqref{eq:ActionScalar} as
\begin{align}
S[\phi,\tilde{j}]=\sum_{n}&\Big[\frac{1}{2}\int_{k,q}\phi_{-n,-k}\big(G^{\textsl{eff}}_{n,kq}\big)^{-1}\phi_{n,q}\\
&+\int_{k}\phi_{n,k}\big(\tilde{j}_{-k}+2ig\rho\sqrt{t}\delta_{n0}\big)\Big],\notag
\end{align}
and we perform the gaussian integral over the fields $\phi$, by completing the square
  \begin{align}
 \label{eq:integralPhi}
\int D\phi D\tilde{j}\,e^{iS[\phi,\tilde{j}]} = \frac{\Pi_{n}\text{Det}[G^{\textsl{eff}}_{n}]^{1/2}}{\Pi_{n}\text{Det}[G^{0}_{n}]^{1/2}}\cdot \int D\tilde{j}\,&\exp\bigg\{-\frac{i}{2}\int_{k,q}\tilde{j}_{-k}\big(\sum_{n}G^{eff}_{n,kq}\big)\tilde{j}_{q}\\
&+i(2ig\rho\sqrt{t})\int_{k,q}\big(G^{\textsl{eff}}_{0,kq}\big)\tilde{j}_{q}+\frac{i}{2}(2ig\rho\sqrt{t})^{2}\int_{k,q}G^{\textsl{eff}}_{0,kq}\bigg\}.\notag
\end{align}
 
The integral over the Lagrange multiplier is also Gaussian:
 \begin{align}
 \label{eq:integralLagrange}
\int D\tilde{j}\,e^{iS[\tilde{j}]} = \frac{ \text{Det}[\sum_{n}G^{\textsl{eff}}_{n}]^{-1/2}}{\text{Det}[\sum_{n}G^{0}_{n}]^{-1/2}}\cdot \exp\bigg\{+\frac{i}{2}\int_{k_{1}\cdot k_{4}}2ig\rho\sqrt{t}\big(G^{\textsl{eff}}_{0,k_{1}k_{2}}\big)\cdot \big(\sum_{n}G^{\textsl{eff}}_{n}\big)^{-1}_{k_{2}k_{3}}\cdot2ig\rho\sqrt{t}\big(G^{\textsl{eff}}_{0,k_{3}k_{4}}\big)\bigg\}.
\end{align}
By collecting the pieces coming from Eqns.~\eqref{eq:FuncInt}, \eqref{eq:integralPhi} and \eqref{eq:integralLagrange}, one finds Eq.~(17) in the main text.
 \end{widetext}

\end{document}